# An Improved Move-To-Front (IMTF) Off-line Algorithm for the List Accessing Problem

Rakesh Mohanty[1], Sasmita Tripathy[2]

[1]Department of Computer Science and Engineering, Indian Institute of Technology, Chennai , India
[2]Department of Computer Science and Engineering, Veer Surendra Sai University of Technology, Burla, Orissa, India
[1]rakesh.iitmphd@gmail.com, [2]su.smita2008@gmail.com

*Abstract-* For the List Accessing Problem, Move-To-Front (MTF) algorithm has been proved to be the best performing online list accessing algorithm till date in the literature[10]. In this paper, we have made a comprehensive analysis of MTF algorithm and developed an Improved-MTF (IMTF) offline algorithm. We have generated two new types of data set and devise a new method of experimental analysis for our proposed algorithm. Our experimental analysis shows that IMTF is performing better than MTF algorithm.

## I. INTRODUCTION

List accessing problem is a self organizing linear search problem. In linear search, we access a set of elements sequentially in a fixed size unsorted linear list from the start of the list. We reorganize the list by performing exchanges after accessing an element so that the access cost for subsequent elements are reduced thereby reducing total access cost. All these tasks are performed by an algorithm called list accessing algorithm.

### A. List Accessing Cost Model

When we access an element in the list, a cost model is used to compute the access cost. For the Standard full cost model by Sleator and Tarjan, the cost is calculated as follows. For accessing the $i^{th}$ element in the list, access cost is i. Immediately after an access, the accessed element can be moved any distance forward in the list without paying any cost. These exchanges cost nothing and are called free exchanges. For any exchange between two adjacent elements in the list, cost is 1. These exchanges are called paid exchanges. Hence total cost in a full cost model is the sum of number of paid exchanges and the access cost. For the partial cost model, the access cost is calculated by the number of comparisons between the accessed element and the elements present before the accessed element in the list. For accessing the $i^{th}$ element of the list, we have to make i-1 comparisons. Hence the access cost for $i^{th}$ element in partial cost model is i-1.

### B. Applications and Motivation

List accessing problem uses the self organizing data structure. Self organizing data structure reorganizes their structure while processing a sequence of operations. The purpose of this reorganization is to guarantee the efficiency of future operations and improve the performance of this data structure. Bentley and McGeoch experimented on VLSI circuit simulation programs. This program spent 5 minutes in a setup phase, most of which was taken up by linear searches through a symbol table. By using a self organizing linear search, the setup time was reduced to about 30 seconds, most of which was not involved with searching through the symbol table. The list accessing problem can be used in maintaining a list of identifiers in a symbol table by a compiler or interpreter, or for resolving collisions in a hash table. Other applications of list accessing algorithms are for computing point maxima and convex hull. The list accessing algorithms are also widely used for data compression.

### C. Related Work

The list accessing problem is of significant theoretical and practical interest for the last four decades. As per our knowledge, study of list accessing techniques was initiated by the pioneering work of McCabe[1] in 1965. He investigated the problem of maintaining a sequential file and developed two algorithms Move-To-Front(MTF) and Transpose. From 1965 to 1985, the list update problem was studied by many researchers [2], [3], [4], [5] under the assumption that a request sequence is generated by a probability distribution. Hester and Hirschberg[6] have provided an extensive survey of average case analysis of list update algorithms. The seminal paper by Sleator and Tarjan [7] in 1985 made the competitive analysis of online algorithms very popular. The first use of randomization and the demonstration of its advantage in the competitive analysis context was done by Borodin, Linial and Saks [8] with respect to metrical task systems in 1985. Bachrach et. al. have provided an extensive theoretical and experimental study of online list accessing algorithms in 2002 [9]. Quite a few of the competitive analysis of the list update problem were carried out without any specific knowledge of the nature of the optimal offline algorithm.

*D. Our Contribution*

In this paper, we investigate the MTF offline algorithm for the list accessing problem. We provide some important observations on properties of MTF algorithm and developed some interesting formula. We explore and define a novel concept where after accessing an element, with a specific condition, we move the accessed element to the front of the list. By using this concept, we proposed an improved MTF offline algorithm. We have generated two new types of data set and devise a new method of experimental analysis for our proposed algorithm. Our experimental analysis shows that IMTF is performing better than MTF algorithm.

*E. Organization of the paper*

This paper is organized as follows. The analysis of MTF algorithm and related results are presented in section II. Section III contains our new proposed algorithm IMTF. Section IV presents our experimental analysis and associated results obtained from the comparison of access cost of MTF and IMTF. Section V provides the concluding remarks.

## II. PRELIMINARIES

Till date many list accessing algorithms have been developed. There are two types of list accessing algorithms- online and offline. In online algorithms, the request sequence is partially known, i.e. we know the current request only and future requests come on the fly. In offline algorithms, we know the whole request sequence in advance. In our study, we have considered the Move-To-Front offline algorithm for the list accessing problem.

**MTF:** After accessing an element in the list, the accessed element is moved to the front of the list, without changing the relative order of other elements in the list.

*A. MTF Algorithm and its Analysis*

Performance of linear search in an unsorted list can be enhanced by using self reorganizing heuristics that attempts to ensure that frequently accessed elements are near the front of the list. It is proved that MTF algorithm is unique optimal online algorithm for the list accessing problem. MTF performs best when the list has a high degree of locality of reference.

We illustrate the MTF algorithm as follows. Let the list configuration be 1, 2, 3 and request sequence be 2, 3, 1, 3, 2. Each time after accessing, the accessed element is moved to the front of the list. So the total access cost for MTF using full cost model is 2+3+3+1+3=13. From our study and analysis of MTF algorithm, we obtained the following interesting observations.

(1) **MTF algorithm gives minimum cost, when all elements of the request sequence are accessed as the 1st element of the list. The total access cost is equal to n, where n is the size of the request sequence.**

We illustrate the above fact as follows. Considering the configuration of the list as 1 2 3 and request sequence of length 5 having the elements 1 1 1 1 1, we obtain the total access cost using MTF algorithm is 5.

(2) **MTF algorithm gives worst cost, when request sequence is in reverse order of that of the list. Here the total access cost is equal to n*l, where l is the size of the list and n is the size of the request sequence.**

*Illustration:* For configuration of the list 1 2 3 and request sequence having elements 3 2 1 3 2, MTF gives the worst access cost. The total access cost is 3*5=15.

(3) **For distinct elements of request sequence, MTF always performs poorly.**

(4) **If the request sequence consists of repetition of $k^{th}$ element of the list n times, then total access cost for MTF is n+k-1.**

*Illustration:* For the list configuration 1 2 3, the total access cost of MTF for request sequence having elements 2 2 2 2 2 is 5+2-1 = 6. Here n=5, k=2.

After analysis of MTF, we observed that moving the accessed element $x_i$ from position i to the front of the list reduces the total access cost if and only if $x_i$ is present within next i-1 distinct elements from the accessed element in the request sequence, otherwise the movement is unnecessary. Using this idea, we proposed a new offline algorithm called Improve Move-To-Front(IMTF).

## III. UNIQUENESS OF OUR APPROACH

We explore and define a new concept based on look-ahead. Using the look-ahead concept, we have developed an Improved MTF (IMTF) algorithm, which is better than the the MTF algorithm.

*A. IMTF Algorithm*

*After accessing an element $x_i$, in $i^{th}$ position in the list, IMTF moves the accessed element to the front of the list if and only if the accessed element $x_i$ is found by looking ahead the next i-1 elements in the request sequence. If $x_i$ is not present within the next i-1 elements in the request sequence, the list* *configuration remains the same without any movement of $x_i$ in the list.*

*Illustration of IMTF :* Let us consider the list configuration 1 2 3 and request sequence 3 2 1 3 2. Here the first element of the request sequence is 3 which is present in the third position in the list, so the access cost for this element is 3. After accessing, it checks next (3-1) elements in the request sequence, whether the element is present or not. Here it is not present, so the configuration of the list remains the same. Next element of the request sequence is 2 which is present in second position in the list, so the access cost for this element is 2. After accessing 2, next (2-1) one element is checked from the accessed element of the request sequence. Here 2 is not present, So the list remains the same. The process is continued till we reach the end of the request sequence. Here the total access cost is 3+2+1+3+2=11. By applying this algorithm, we restrict the unnecessary movement of the accessed element to reduce the access cost for future elements of the request sequence and the rearrangement cost of the list.

## IV. EXPERIMENTAL ANALYSIS

We have performed two experiments by implementing MTF and IMTF algorithms. For each algorithm, we have generated the list and request sequence for different data set. We have calculated the total access cost of each algorithm for different request sequence size and list size.

### A. Data Set Generation

The inputs for the list accessing algorithm are a list and a request sequence. We generated two different types of request sequence and corresponding list based on the nature of the data set. The first data set consists of alphabetic and special character set. The second data set consists of numerical data using different base number system such as decimal, binary, octal and hexadecimal.

*1) Alphabetic And Special Character Data Set*

A linear linked list is generated from randomly selected characters from the key board, by considering each character as a node in the linked list. The random characters are selected from different text files. We perform experiments by varying the size of request sequence from 100 to 1000. These types of request sequence usually appear in sms writing, report writing etc. Another linear linked list is generated from input request sequence. Each distinct character of request sequence is considered as a node in the linked list. Here maximum size of the list is 92 because a file contains maximum 92 distinct characters. It includes all 52 alphabets of English language (both upper case and lower case), 30 special characters and 10 digits. This data set generated is used in our first experiment to produce TABLE-1 results.

*2) Numerical Data Set in Different Number Systems*

In this experiment, a linear linked list is generated for request sequence by randomly selecting digits from the key board. Here the base value corresponds to the maximum list size. We conduct the experiment for binary, octal, decimal and hexadecimal number by varying the maximum list size as 2, 8, 10 and 16 respectively. This data set generated is used in our second experiment to produce TABLE-2 results.

## V. IMPLIMENTATION RESULTS

In our experiment, we implemented both the MTF and IMTF algorithms in C language with Windows XP operating system. Here we use the singly linked list data structure. In our source code, we declare a structure data type for generating two singly linked lists one for request sequence and the other for the list. In our code we use following user defined functions. The function List () generates a Linked list for input list and stores the distinct character of request sequence in the information part of the node of the linked list. REQ () function is used to generate another linear linked list for request sequence and stores all the character of request sequence in the information part of the linked list. MTF () function is used for calculating the total access cost using MTF algorithm for a given list and a request sequence. This function contains two loops -the outer loop is for the request sequence and the inner loops for the input list. Each time a character is read from the request sequence and that element is searched in the list by traversing the linked list generated. The position of requested element in the list is the access cost for that element. That cost is stored in a variable. After finding the requested element in the list, it is deleted from its original position and inserted in the front of linked list. This process is continued till end the outer loop. Outside the outer loop display the total access cost for all element of request sequence. IMTF () function is used for calculating the total access cost using IMTF algorithm for same list and request sequence. Each time a character is read from the request sequence and that element is searched in the input list by traversing the linked list generated. After accessing the element, find the position (i) of accessed element in the list and check next i-1 elements from the current requested element in the request sequence if the accessed element present within the next i-1 elements of the request sequence then call MTF () algorithm else return from the IMTF(). We conduct different experiments with changing the list configuration and request sequence of different sizes and calculate the total access cost for MTF and IMTF algorithms. After finding the total access cost of MTF and IMTF algorithms, we calculate the gain in IMTF algorithm over MTF as follows. Let L be the size of the list

*An Improved Move-To-Front (IMTF) Off-line Algorithm for the List Accessing Problem*

and N be the size of the request sequence. Let $C_{MTF}$ be the total access cost of MTF algorithm. Let $C_{IMTF}$ be the total access cost of IMTF algorithms. Then gain (g) is defined as follows.

$$g = ((C_{IMTF} - C_{MTF})/C_{MTF}) * 100.$$

**TABLE 1- Experimental Results with Alphabetic and special character data set.**

| N | L | $C_{MTF}$ | $C_{IMTF}$ | g |
|---|---|---|---|---|
| 100 | 23 | 867 | **702** | 19.03% |
| 00 | 26 | 1954 | **1456** | 25.38% |
| 300 | 26 | 3287 | **2542** | 22.66% |
| 400 | 26 | 4403 | **3377** | 23.30% |
| 500 | 45 | 12724 | **8851** | 30.57% |
| 600 | 26 | 6701 | **5106** | 23.80% |
| 700 | 52 | 10519 | **8369** | 20.43% |
| 800 | 65 | 15671 | **1879** | 24.19% |
| 900 | 70 | 19692 | **14533** | 25.19% |
| 1000 | 70 | 23591 | **15277** | 35.24% |

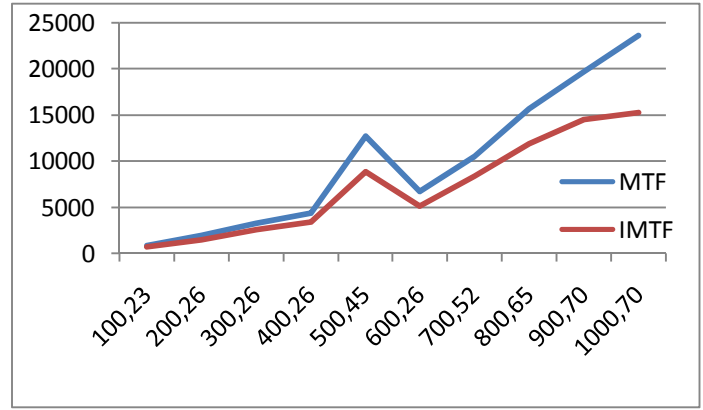

Line chart for Alphabetic and Special Character Data Set

**TABLE 2 -Experimental Results with Numerical data set with Hexadecimal, Decimal, Octal and Binary**

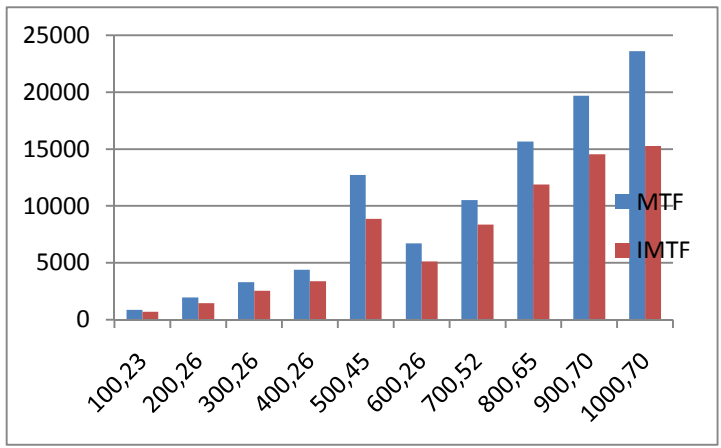

Bar chart for Alphabetic and Special Character Data Set

| N | L | C$_{MTF}$ | **C$_{IMTF}$** | G |
|---|---|---|---|---|
| 50 | 16 | 492 | **373** | 24.18% |
| 50 | 10 | 325 | **234** | 28% |
| 50 | 8 | 268 | **206** | 23.13% |
| 50 | 2 | 76 | **67** | 11.84% |
| 100 | 16 | 970 | **708** | 27.01% |
| 100 | 10 | 632 | **481** | 23.89% |
| 100 | 8 | 477 | **387** | 16.86% |
| 100 | 2 | 161 | **140** | 13.04% |
| 200 | 16 | 1570 | **1330** | 15.28% |
| 200 | 10 | 1177 | **858** | 27.10% |
| 200 | 8 | 983 | **805** | 18.10% |
| 200 | 2 | 316 | **280** | 11.39% |

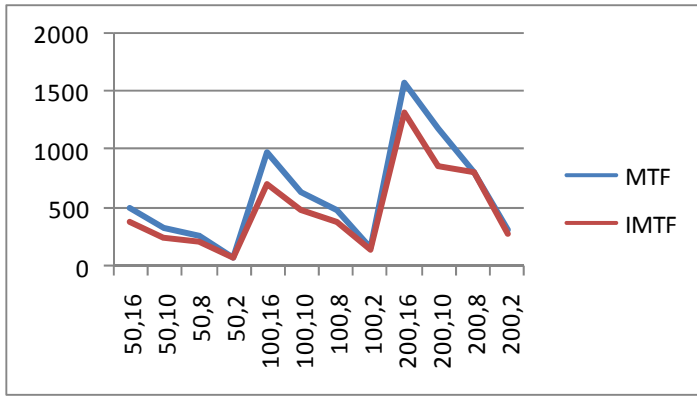

Line chart for Numerical data set with Hexadecimal, Decimal, Octal and Binary Number

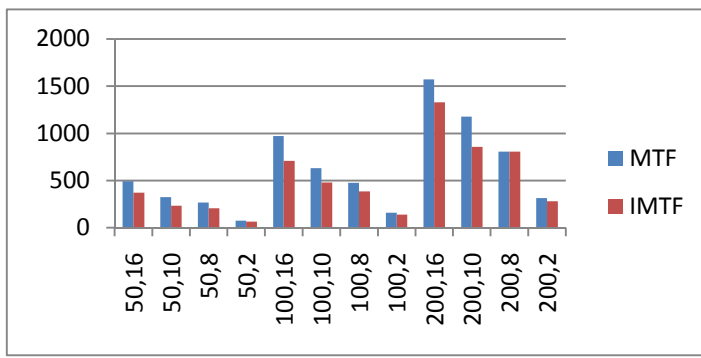

Bar chart for Numerical data set with Hexadecimal, Decimal, Octal and Binary Number

## VI. CONCLUDING REMARKS

In this paper, we have presented an Improved Move-To-Front (IMTF) offline algorithm for the list update problem, which is observed to be performing better than MTF algorithm. A comparative performance analysis of IMTF over MTF has been done. There is significant performance gain of IMTF over MTF algorithm. The performance gain has been observed to be more than 10 for smaller size lists and more than 20 for larger size lists.

## VII. ACKNOWLEDGEMENT

Our special thanks to Dr N. S. Narayanaswamy of Department of Computer Science and Engineering of Indian Institute of Technology Madras for his initial motivation and extensive support for the carrying out this research work.

## Authors Biography

Prof. Rakesh Mohanty is currenty working as a faculty member in Department of Computer Science and Engineering in Veer Surendra Sai University of Technology, Burla, Sambalpur, Orissa, India. He obtained his B.E. and M.Tech. in Computer Science and Engineering from University College of Engineering, Burla and Jawaharlal Nehru University, Newdelhi in 1998 and 2002 respectively. Currently he is pursuing his PhD in Computer Science and Engineering in Indian Institute of Technology Madras, Chennai, India. His areas of research interests are Data Structures, Algorithms, Scheduling and Routing.

Sasmita Tripathy has completed her M.Tech. in July 2010 from Department of Computer Science and Engineering, Veer Surendra Sai University of Technology, Burla, Sambalpur, Orissa, India. She obtained her Master of Computer Application degree from Biju Patnaik University of Technology, Rourkela, Orissa in 2005. Currently she is teaching as a faculty member in MCA Department of Veer Surendra Sai University of Technology. Her areas of research interests are Data Structures and Algorithms.